\documentclass{camera}
\usepackage{graphicx} 

\begin{document}

\newcommand{\lsim}{\mathrel{\mathop{\kern 0pt \rlap
  {\raise.2ex\hbox{$<$}}}
  \lower.9ex\hbox{\kern-.190em $\sim$}}}
\newcommand{\gsim}{\mathrel{\mathop{\kern 0pt \rlap
  {\raise.2ex\hbox{$>$}}}
  \lower.9ex\hbox{\kern-.190em $\sim$}}}

\title{Particle Dark Matter in DAMA/LIBRA}

\author{R. Bernabei$^a$, P. Belli$^a$, F. Cappella$^b$, R. Cerulli$^c$,\\
        C.J. Dai$^d$, A. d'Angelo$^b$, H.L. He$^d$, A. Incicchitti$^b$,\\ 
        X.M. Ma$^d$, F. Montecchia$^{a,e}$, F. Nozzoli$^a$,\\ 
        D. Prosperi$^{b,}$\footnote{Deceased}, X.D. Sheng$^d$, Z.P. Ye$^{d,f}$ and R.G. Wang$^d$}
\vspace{-2.cm}
\organization{$^a$Dip. di Fisica, Universit\`a di Roma ``Tor Vergata'' and \\
                  INFN, sez. Roma ``Tor Vergata'', I-00133 Rome, Italy\\
              $^b$Dip. di Fisica, Universit\`a di Roma ``La Sapienza'' and \\
                  INFN, sez. Roma, I-00185 Rome, Italy\\
              $^c$Laboratori Nazionali del Gran Sasso, I.N.F.N., Assergi, Italy\\
              $^d$IHEP, Chinese Academy, P.O. Box 918/3, Beijing 100039, China\\
              $^e$also: Lab. Sperim. Policentrico di Ingegneria Medica, Universit\`a di Roma Tor Vergata\\
              $^f$also: University of Jing Gangshan, Jiangxi, China}

\maketitle
\vspace{-1.cm}
\begin{abstract}
The present DAMA/LIBRA experiment 
and the former DAMA/NaI have cumulatively released so far 
the results obtained with the data collected over 13 annual cycles (total exposure: 
1.17 ton $\times$ yr). They give a model 
independent evidence of the presence of DM particles in the galactic halo on the basis
of the DM annual modulation signature at 8.9 $\sigma$ C.L. for the cumulative exposure.
\end{abstract}

\section{Introduction}

The DAMA project is based on the development and use of low background scintillators, and 
several low background set-ups have been realized and used for various kinds of investigations
\cite{damaweb}.
In particular, the former DAMA/NaI and the present DAMA/LIBRA 
experiments at the Gran Sasso National Laboratory of the INFN have the main aim to investigate the 
presence of Dark Matter particles in the galactic halo by exploiting the model independent 
DM annual modulation signature, originally suggested in the mid 80's 
in ref. \cite{Freese}. 
In fact, as a consequence of its annual revolution around the Sun, which is moving in the Galaxy
travelling with respect to the Local Standard of Rest towards 
the star Vega near the constellation of Hercules, 
the Earth should be crossed by a larger flux of Dark Matter particles around $\sim$ 2 June
(when the Earth orbital velocity is summed to the one of the solar system with respect 
to the Galaxy) and by a smaller one around $\sim$ 2 December (when the two velocities are subtracted).
It is worth noting that this signature has a different origin and peculiarities than the seasons on the Earth
and than effects correlated with seasons (consider the expected value of the 
phase as well as the other requirements listed below). 
This annual modulation signature is very distinctive since the effect induced by DM
particles must simultaneously satisfy all the following requirements:
the rate must contain a component
modulated according to a cosine function (1) with one year period (2)
and a phase that peaks roughly around $\simeq$ 2$^{nd}$ June (3);
this modulation must only be found
in a well-defined low energy range, where DM particle induced events
can be present (4); it must apply only to those events in
which just one detector of many actually ``fires'' ({\it single-hit} events), since
the DM particle multi-interaction probability is negligible (5); the modulation
amplitude in the region of maximal sensitivity must be $\lsim$7$\%$
for usually adopted halo distributions (6), but it can
be larger in case of some possible scenarios such as e.g. those in refs. \cite{Wei01,Fre04}. 
This offers an efficient DM model independent signature, able to test a large interval of
cross sections and of halo densities; moreover, the use of highly 
radiopure NaI(Tl) scintillators as target-detectors assures sensitivity to wide 
ranges of DM candidates, of interaction types and of astrophysical scenarios.

It is worth noting that only systematic effects or side reactions 
able to simultaneously fulfil all the requirements given above (and no one has ever been
suggested over more than a decade) and to account for the whole observed modulation 
amplitude might mimic this DM signature.

The description, radiopurity and main
features of the DAMA/LIBRA set-up
are discussed in details in ref. \cite{perflibra,modlibra}; moreover, this set-up has firstly been 
upgraded 
in September/October 2008 \cite{modlibra2}. 

The cumulative DA\-MA/LI\-BRA exposure -- after the new data release occurred at beginning of 
2010 \cite{modlibra2} -- 
is 0.87 ton $\times$ yr (6 annual cycles), and cumulatively with DA\-MA/NaI the 
exposure is 1.17 ton $\times$ yr (13 annual cycles in total).

\section{The model independent result}

Several analyses on the model-independent investigation of the DM annual 
modulation signature have been performed in \cite{modlibra2} as previously done 
\begin{figure}[!ht]
\begin{center}
\vspace{-0.4cm}
\includegraphics[width=11.5cm] {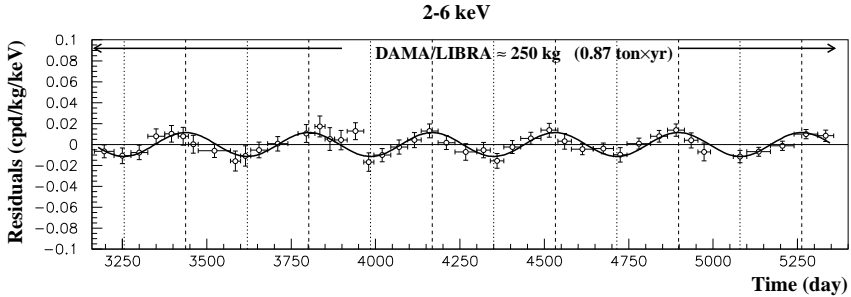}
\end{center}
\vspace{-0.8cm}
\caption{Experimental model-independent residual rate of the {\it single-hit} scintillation events, 
measured by DAMA/LIBRA-1,2,3,4,5,6 in the (2 -- 6) keV energy interval
as a function of the time \cite{modlibra,modlibra2}. The zero of the time scale is January 1$^{st}$ 
of the first year of data taking of the former DAMA/NaI experiment. 
The experimental points present the errors as vertical bars and the associated time bin width as horizontal bars. 
The superimposed curve is the cosinusoidal function behavior $A \cos \omega(t-t_0)$  
with a period $T = \frac{2\pi}{\omega} =  1$ yr, with a phase $t_0 = 152.5$ day (June 2$^{nd}$) and with
modulation amplitude, $A$, equal to the central value obtained by best fit over the whole data 
including also the exposure previously collected by the former DAMA/NaI experiment.
The dashed vertical lines 
correspond to the maximum expected for the DM signal (June 2$^{nd}$), while 
the dotted vertical lines correspond to the minimum. See refs. \cite{modlibra,modlibra2} and refs. therein.}
\label{fg:res}
\end{figure}
in ref. \cite{modlibra} and refs. therein. In particular, Fig. \ref{fg:res} shows 
the time behaviour 
of the experimental residual rates for {\it single-hit} 
events in the (2--6) keV energy interval; as known, 
here and hereafter keV means keV electron equivalent. 
The hypothesis of absence of modulation in the data can be discarded \cite{modlibra,modlibra2}. 
Moreover, when the period and the phase parameters as well as the modulation amplitude are 
kept free fitting the experimental residuals of Fig. \ref{fg:res} 
with the formula: Acos$\omega$(t - t$_0$),
values well compatible with the expectations for a signal 
in the DM annual modulation signature are found \cite{modlibra,modlibra2}.
In particular, the phase -- whose better 
determination is obtained by using a maximum likelihood 
analysis \cite{modlibra,modlibra2} -- is consistent with about June $2^{nd}$ within $2\sigma$.
For completeness, we note that a slight energy dependence of the phase 
could be expected in case of possible contributions of non-thermalized DM components 
to the galactic halo, such as e.g. the SagDEG stream \cite{epj06} 
and the caustics \cite{caus}.

The data have also been 
investigated by a Fourier analysis, obtaining a clear peak corresponding to a period of 1 year; 
the same analysis in other energy region shows instead only aliasing peaks 
\cite{modlibra,modlibra2}.

The measured energy distribution has been investigated in
other energy regions not of interest for Dark Matter,
also verifying the absence of any significant background modulation\footnote{In fact, 
the background in the lowest energy region is
essentially due to ``Compton'' electrons, X-rays and/or Auger
electrons, muon induced events, etc., which are strictly correlated
with the events in the higher energy part of the spectrum.
Thus, if a modulation detected
in the lowest energy region would be due to
a modulation of the background (rather than to a signal),
an equal or larger 
modulation in the higher energy regions should be present.}. 
In particular, the measured rate
integrated above 90 keV, R$_{90}$, as a function of the time has been analysed;
fitting its time behaviour 
with phase and period as for DM particles, a modulation amplitude compatible with zero
is found excluding the presence of any background
modulation in the whole energy spectrum at a level much
lower than the effect found in the lowest energy region for the {\it single-hit} events 
\cite{modlibra,modlibra2}.
Similar result is obtained when comparing 
the {\it single-hit} residuals in the (2--6) keV with those 
in other energy intervals \cite{modlibra,modlibra2}.
It is worth noting that the obtained results already account for whatever 
kind of background and, in addition, that no background process able to mimic
the DM annual modulation signature (that is able to simultaneously satisfy 
all the peculiarities of the signature and to account for the measured modulation amplitude)
is available (see also discussions e.g. in \cite{modlibra,scineghe09}, 
refs. therein and later).

\begin{figure}[!ht]
\vspace{-0.4cm}
\begin{center}
\includegraphics[width=11.5cm] {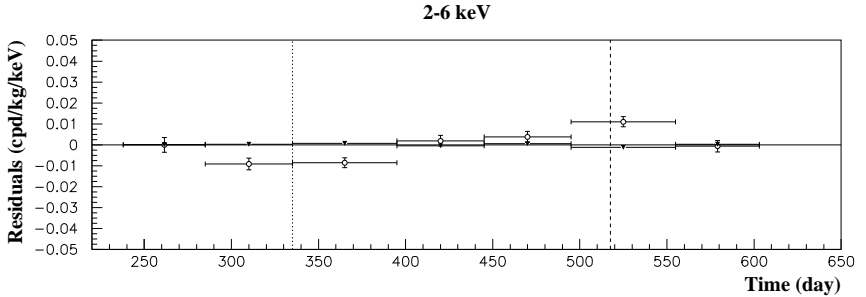}
\end{center}
\vspace{-0.8cm}
\caption{Experimental residual rates over the six DAMA/LIBRA annual cycles for {\it single-hit} events 
(open 
circles) (class of events to which DM events belong) and for {\it multiple-hit} events (filled triangles)
(class of events to which DM events do not belong).
They have been obtained by considering for each class of events the data as collected in a 
single annual cycle 
and by using in both cases the same identical hardware and the same identical software procedures.
The initial time of the figure is taken on August 7$^{th}$.
The experimental points present the errors as vertical bars and the associated time bin width as horizontal 
bars. Analogous results were obtained for the DAMA/NaI data \cite{ijmd}.
See refs. \cite{modlibra,modlibra2}.}
\label{fig_mul}
\end{figure}

A further relevant investigation has been performed by applying the same hardware and software 
procedures, used to acquire and to analyse the {\it single-hit} residual rate, to the 
{\it multiple-hit} one. 
In fact, since the probability that a DM particle interacts in more than one detector 
is negligible, a DM signal can be present just in the {\it single-hit} residual rate.
Thus, the comparison of the results of the {\it single-hit} events with those of the  {\it 
multiple-hit} ones corresponds practically to compare between them the cases of DM particles beam-on 
and beam-off.
This procedure also allows an additional test of the background behaviour in the same energy interval 
where the positive effect is observed. 
In particular, in Fig. \ref{fig_mul} the residual rates of the {\it single-hit} events measured over 
the six DAMA/LIBRA annual
cycles are reported, as collected in a single cycle, together with the residual rates 
of the {\it multiple-hit} events, in the considered energy interval.
While, as already observed, a clear modulation, satisfying all the peculiarities of the DM
annual modulation signature, is present in 
the {\it single-hit} events,
the fitted modulation amplitude for the {\it multiple-hit}
residual rate is well compatible with zero \cite{modlibra2}.
Thus, again evidence of annual modulation with proper features as required by the DM annual 
modulation signature is present in the {\it single-hit} residuals (events class to which the
DM particle induced events belong), while it is absent in the {\it multiple-hit} residual 
rate (event class to which only background events belong).
Since the same identical hardware and the same identical software procedures have been used to 
analyse the
two classes of events, the obtained result offers an additional strong support for the 
presence of a DM
particle component in the galactic halo.

\begin{figure}[!ht]
\begin{center}
\includegraphics[width=11.5cm] {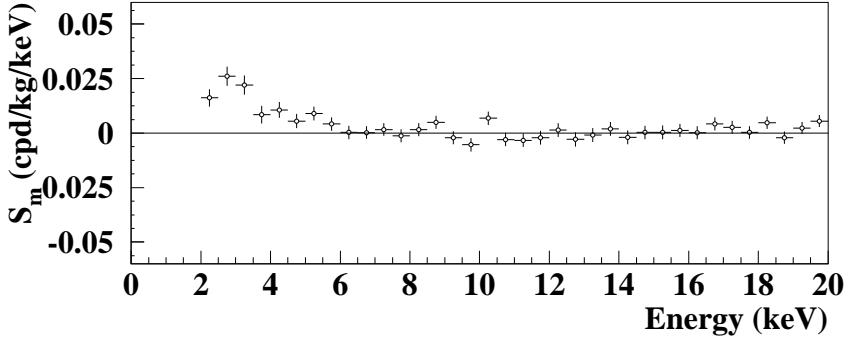}
\end{center}
\vspace{-0.8cm}
\caption{Energy distribution of $S_{m}$ for the
total cumulative exposure 1.17 ton$\times$yr. The energy bin is 0.5 keV.
A clear modulation is present in the lowest energy region,
while $S_{m}$ values compatible with zero are present just above. In fact, the $S_{m}$ values
in the (6--20) keV energy interval have random fluctuations around zero with
$\chi^2$ equal to 27.5 for 28 degrees of freedom.
See refs. \cite{modlibra,modlibra2}.}
\label{sme}
\end{figure}

The annual modulation present at low energy can also 
be shown by depicting -- as a function of the energy --
the modulation amplitude, $S_{m}$, obtained
by maximum likelihood method over the data considering $T=$1 yr and 
$t_0=$ 152.5 day,
as described in refs. \cite{modlibra,modlibra2}.
In Fig. \ref{sme} the obtained $S_{m}$  are shown 
in each considered energy bin (there $\Delta E = 0.5$ keV).
It can be inferred that positive signal is present in the (2--6) keV energy interval, while $S_{m}$
values compatible with zero are present just above. In fact, the $S_{m}$ values
in the (6--20) keV energy interval have random fluctuations around zero with
$\chi^2$ equal to 27.5 for 28 degrees of freedom.
All this confirms the previous analyses.

The method also allows the extraction of the the $S_{m}$ 
values for each detector, for each annual cycle and 
for each energy bin. Thus, following the procedure described in ref. \cite{modlibra},
we have also verified that the S$_{m}$ 
are statistically well distributed in all the annual cycles
and energy bins for each detector \cite{modlibra,modlibra2}.
Among further additional tests, the analysis 
of the
modulation amplitudes as a function of the energy separately for
the nine inner detectors and the remaining external ones has been carried out;
the values are fully in 
agreement, showing that the
effect is well shared between inner and external detectors \cite{modlibra,modlibra2}. 

Finally, releasing the assumption of a phase $t_0=152.5$ day in the maximum likelihood procedure
to evaluate the modulation amplitudes from the data of the 1.17 ton $\times$ yr exposure,
one can alternatively write the signal as \cite{modlibra,modlibra2}:
$ S_{ik} = S_{0,k} + S_{m,k} \cos\omega(t_i-t_0)+Z_{m,k}\sin\omega(t_i-t_0) 
= S_{0,k}+Y_{m,k}\cos\omega(t_i-t^*)$. For signals induced by DM particles one would expect: 
i) $Z_{m,k} \sim 0$ (because of the orthogonality between the cosine and the sine functions); 
ii) $S_{m,k} \simeq Y_{m,k}$; iii) $t^* \simeq t_0=152.5$ day. 
These conditions hold for most of the dark halo models; however, as mentioned above,
slight differences can be expected in case of possible contributions
from non-thermalized DM components, such as e.g. the SagDEG stream \cite{epj06} 
and the caustics \cite{caus}.

\begin{figure}[!ht]
\begin{center}
\includegraphics[width=6.cm] {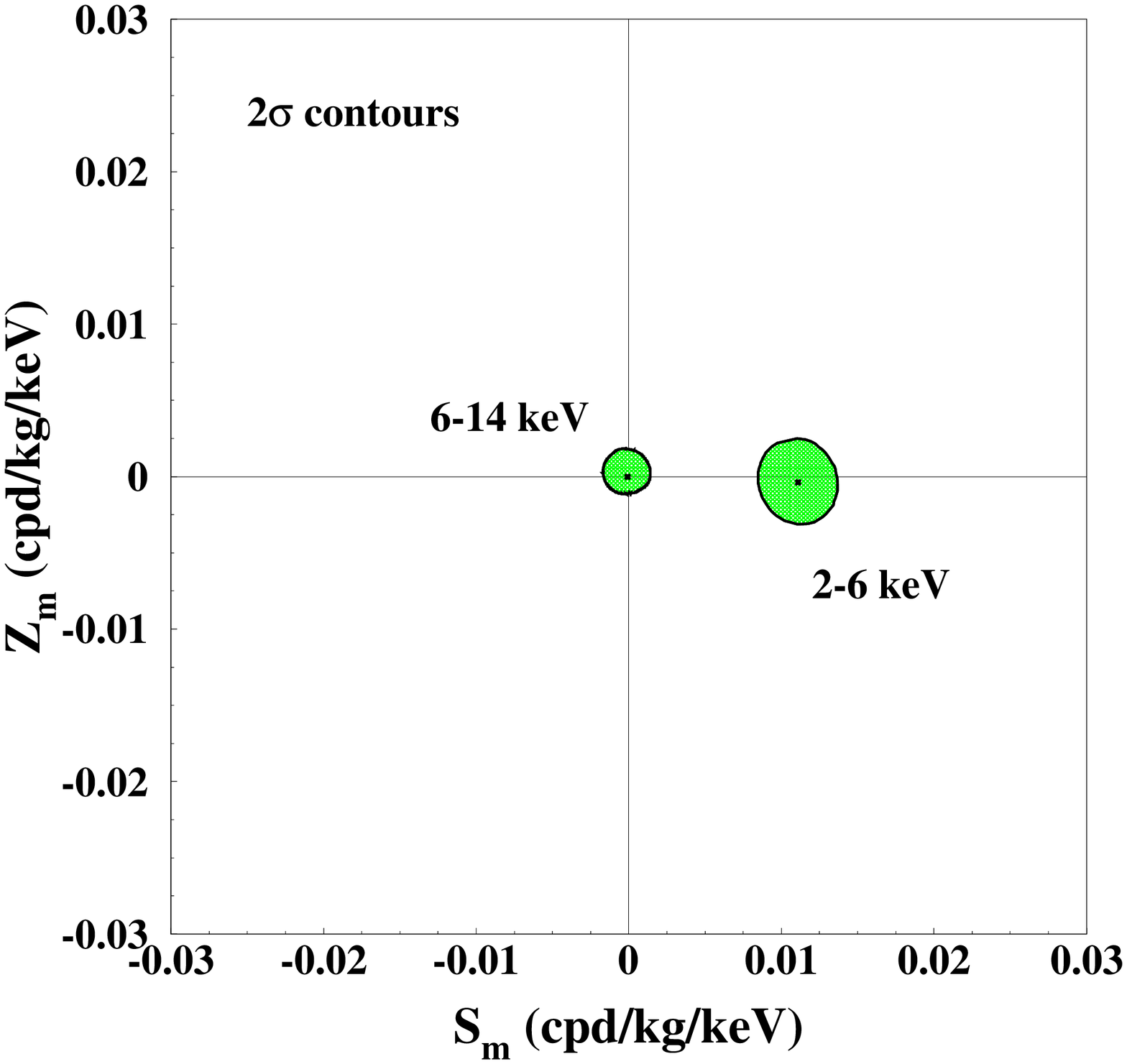}
\includegraphics[width=6.cm] {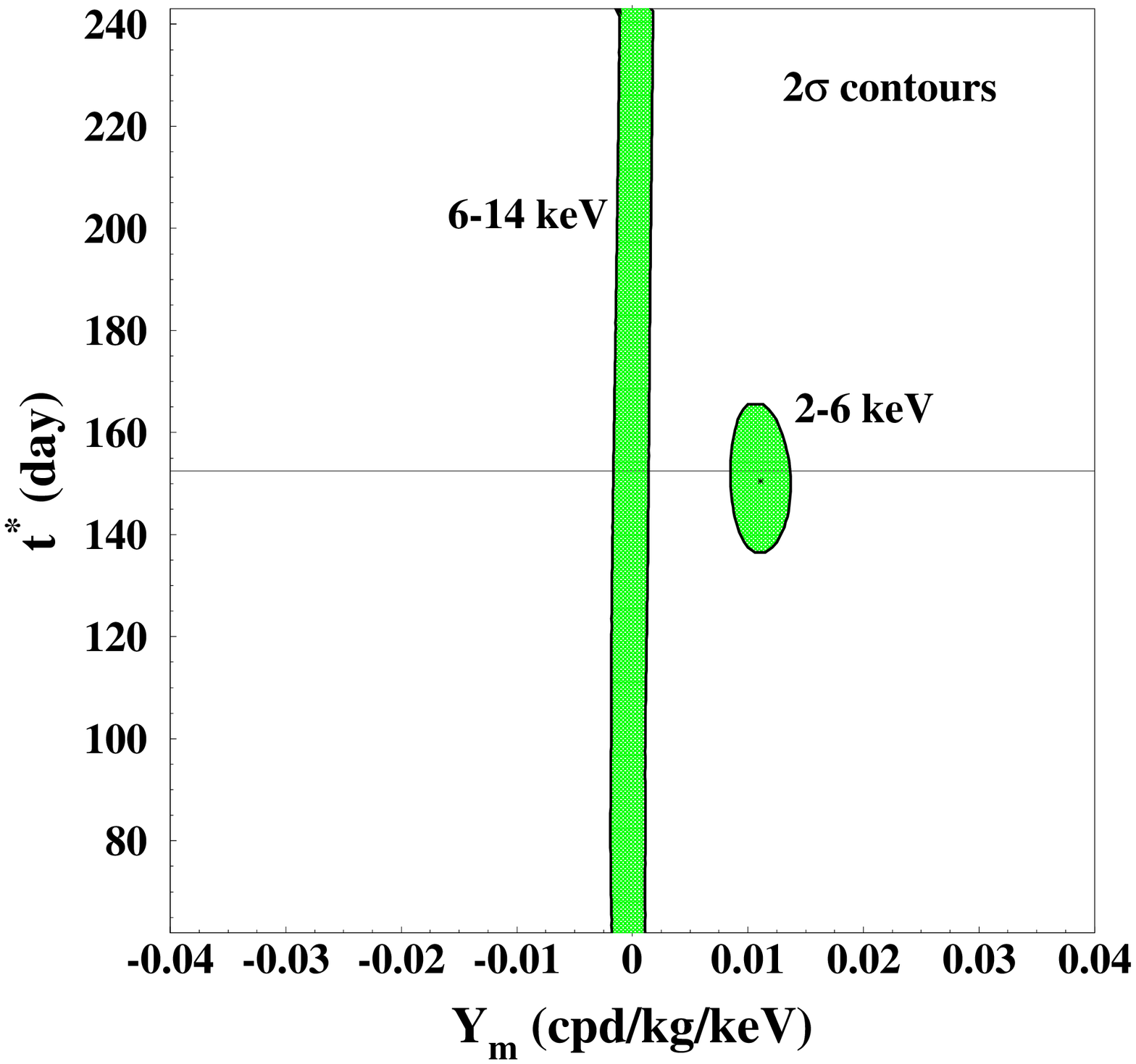}
\end{center}
\vspace{-0.8cm}
\caption{$2\sigma$ contours in the plane $(S_m , Z_m)$ ({\it left})
and in the plane $(Y_m , t^*)$ ({\it right})
for the (2--6) keV and (6--14) keV energy intervals.
The contours have been  
obtained by the maximum likelihood method, considering 
the cumulative exposure of 1.17 ton $\times$ yr.
A modulation amplitude is present in the lower energy intervals 
and the phase agrees with that expected for DM induced signals.
See refs. \cite{modlibra,modlibra2}.}
\label{fg:bid}
\end{figure}

Fig. \ref{fg:bid} shows the 
$2\sigma$ contours in the planes $(S_m , Z_m)$ and $(Y_m , t^*)$
for the (2--6) keV and (6--14) keV energy intervals.
The best fit values are reported in \cite{modlibra2}.
Then, forcing to zero the contribution of the cosine function,
the $Z_{m}$ values as function of the energy have also been determined
by using the same procedure. The values of $Z_{m}$ are expected to be zero;
by the fact, the $\chi^2$ test supports such a hypothesis \cite{modlibra2}.
As in the previous analyses, an annual modulation effect is present in the {\it single-hit}
events in the lower energy interval
and the phase agrees with that expected for DM induced signals.
These results confirm those achieved by other kinds of analyses.

We further stress that sometimes naive statements were put forwards as the fact that
in nature several phenomena may show some kind of periodicity.
It is worth noting that the point is whether they might
mimic the annual modulation signature in DAMA/LIBRA (and former DAMA/NaI), i.e. whether they
might be not only quantitatively able to account for the observed
modulation amplitude but also able to contemporaneously
satisfy all the requirements of the DM annual modulation signature. The same is also for side reactions.
This has already been deeply investigated e.g. in ref. \cite{modlibra,perflibra,modlibra2} and references 
therein. Some additional arguments have also been recently addressed in 
\cite{scineghe09,taupnoz}. See also later.

No modulation has been found in any  
possible source of systematics or side reactions for DAMA/LIBRA as well;
moreover, no one is able to mimic the signature. Thus, cautious upper limits 
(90\% C.L.) on the possible contributions to the DAMA/LIBRA measured modulation amplitude
have been estimated and are summarized e.g. in Table of ref. \cite{modlibra}.

Just as an example we recall here the case of muons, whose flux has been reported
by the MACRO experiment to have an about 2\% modulation with phase around 
mid--July \cite{Mac97}; recently, also LVD and Borexino results \cite{LVD,borexino} 
have been reported. We have already demonstrated that 
not only this effect would give rise in the DAMA set-ups to a quantitatively
negligible contribution (see e.g. \cite{modlibra,modlibra2,RNC,ijmd} and refs. therein), 
but several of the six requirements necessary to mimic
the DM annual modulation signature would also fail: e.g.
muon would also induce modulation in the {\it multiple-hit} events and in the whole energy distribution,
which is not observed. Moreover, even the 
pessimistic assumption of whatever hypothetical 
(even exotic) possible cosmogenic
product -- whose decay or de-ex\-ci\-ta\-tion or whatever else 
might produce: i) only events at low energy; ii) only {\it single-hit}
events; iii) no sizeable effect in the {\it multiple-hits} counting rate --
cannot give rise to any side process able to mimic the investigated DM signature;
in fact, not only this latter hypothetical process
would be quantitatively negligible (see e.g. \cite{modlibra,modlibra2} and refs. therein), but 
in addition its phase 
would be (much) larger than July 15th, and therefore well
different from the one measured by the DAMA experiments and expected
by the DM annual modulation signature ($\simeq$ June 2nd). 
In addition, the phase measured by DAMA experiments in each annual cycle is always around June 2nd, while
the muon phase varies from year to year depending on the condition in the atmosphere.
In particular, the value $(146 \pm 7)$ days \cite{modlibra2} 
measured by DAMA is 5.6 $\sigma$ distant from the LVD mean value ($\simeq 185$ days)
and even more from the 
MACRO one; similar conclusions also hold for Borexino \cite{borexino}.
To be clear, let us note that even a phase value +3$\sigma$ from the one measured for the DAMA observed effect -- that is  
mid-June -- cannot match the muon phase unless for one exceptional year; in fact, 
the maximum outer atmosphere temperature variation 
(and consequently the atmosphere density variation which causes the muon flux modulation)
is typically not in that period at Gran Sasso location.
In conclusion, any possible effect from muons is safely excluded on the basis of all the 
given quantitative facts (and just one of them is enough).

There has been suggested in a recent paper \cite{ralston}
that environmental neutrons (mainly thermal and/or epithermal) once captured by Iodine, might be
responsible of the observed modulation through $^{128}$I decay, 
that produces -- among others -- low energy X-rays/Auger electrons,
provided any hypothetical modulation of the neutron impinging the sensitive part of the detectors
inside the multi-component multi-ton shield. 
To be as clear as possible, we skip several comments about 
that paper and we just summarize few points.
Firstly, we stress that 
-- as already quantitatively discussed e.g. in \cite{RNC,ijmd,modlibra,modlibra2} -- 
environmental neutrons
cannot give any significant contribution to the annual modulation measured by the DAMA experiments
\footnote{For completeness, we also recall that the experimental set-ups, located deep underground,
were equipped with a neutron shield made by Cd foils and
polyethylene/paraffin moderator; moreover, a $\simeq$ 1 m concrete almost completely
surrounds the installation acting as a further neutron moderator. The effectiveness of this shield 
has also been demonstrated e.g. in ref. \cite{modlibra} where a reduction  
larger than one order of magnitude has been measured for the thermal neutrons.} and that $^{128}$I 
-- if any -- cannot mimic the DM annual modulation signature since some of its peculiarities would fail.
Moreover, when the $^{128}$I decays in the EC 
channel, it produces low energy X-rays and Auger electrons,
but -- since the $^{128}$I would be inside the NaI(Tl) detectors -- the
detectors would measure the total energy
release of all the X-rays and Auger electrons emitted following the EC,
that is the atomic binding
energy either of the $K$ shell (32 keV) or of the $L$ shells (4.3 to 5 keV)
of the $^{128}$Te. The probability that so low-energy gamma's and electrons would escape a detector
is very small; thus, we can conclude that:
1) the $L$-shell contribution would be a gaussian centered around 4.5 keV; but this is excluded 
by the DAMA data (see Fig. \ref{sme}). Moreover, 
the efficiencies to detect an event per one $^{128}$I decay are:
2 $\times$ 10$^{-3}$, 6 $\times$ 10$^{-3}$, and 2 $\times$ 10$^{-3}$ 
in (2--4) keV, (4--6) keV and (6--8) keV 
respectively, as calculated by Montecarlo code.
Thus, in addition, the contribution of $^{128}$I in the (2--4) keV -- if any --
would be similar to the one in the (6--8) keV, while the data exclude that;
2) the $K$-shell contribution (around 30 keV) must be 8 times larger than
that of $L$-shell, while no modulation has been observed above 6 keV and, in particular, around
30 keV;
3) the $^{128}$I also decays by $\beta^-$ with much larger branching ratio than EC and with $\beta^-$
end-point energy at 2 MeV. Again, no modulation has instead been observed in DAMA experiments 
at high energy \cite{modlibra,modlibra2}.
\begin{figure}[!ht]
\begin{center}
\includegraphics[width=9.cm] {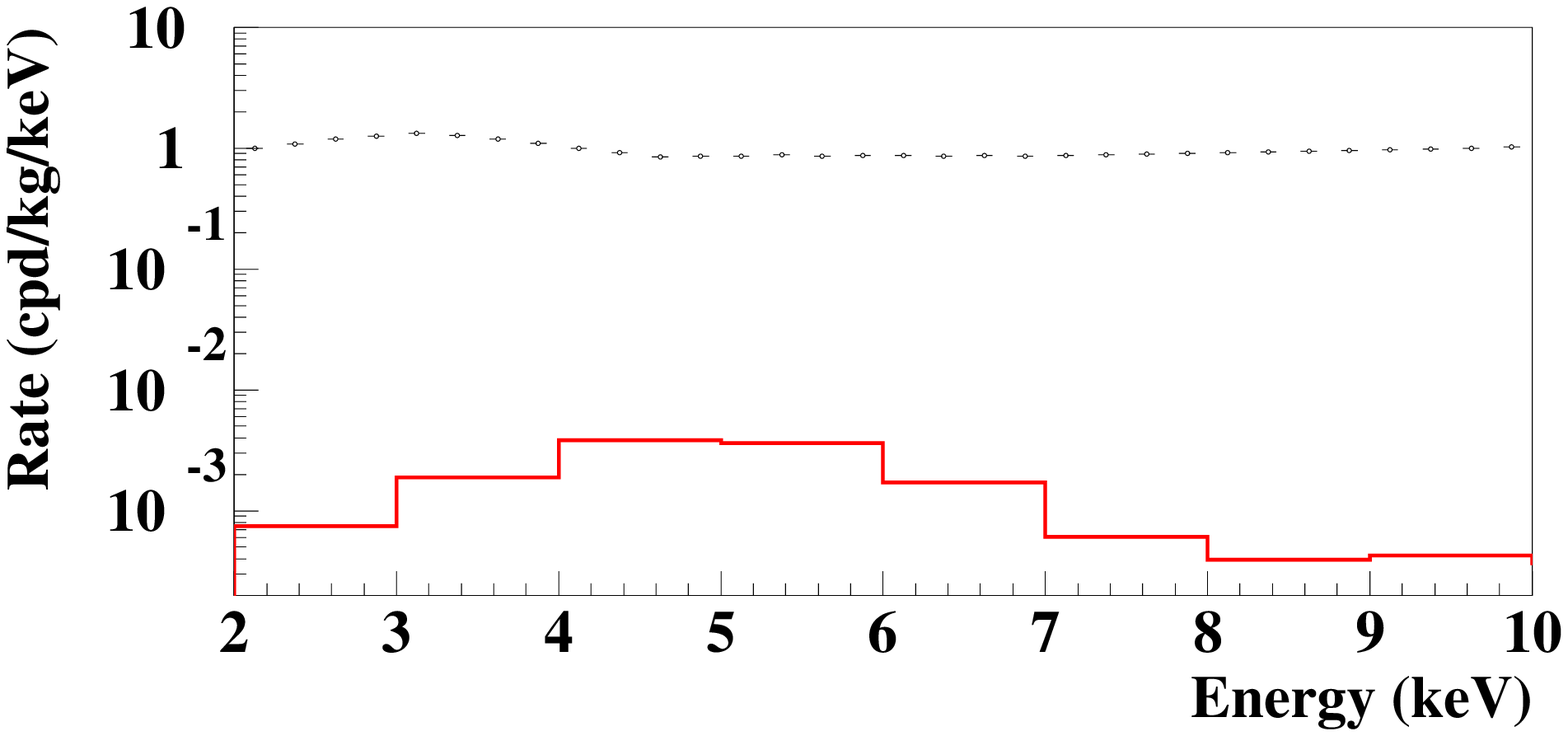}
\includegraphics[width=9.cm] {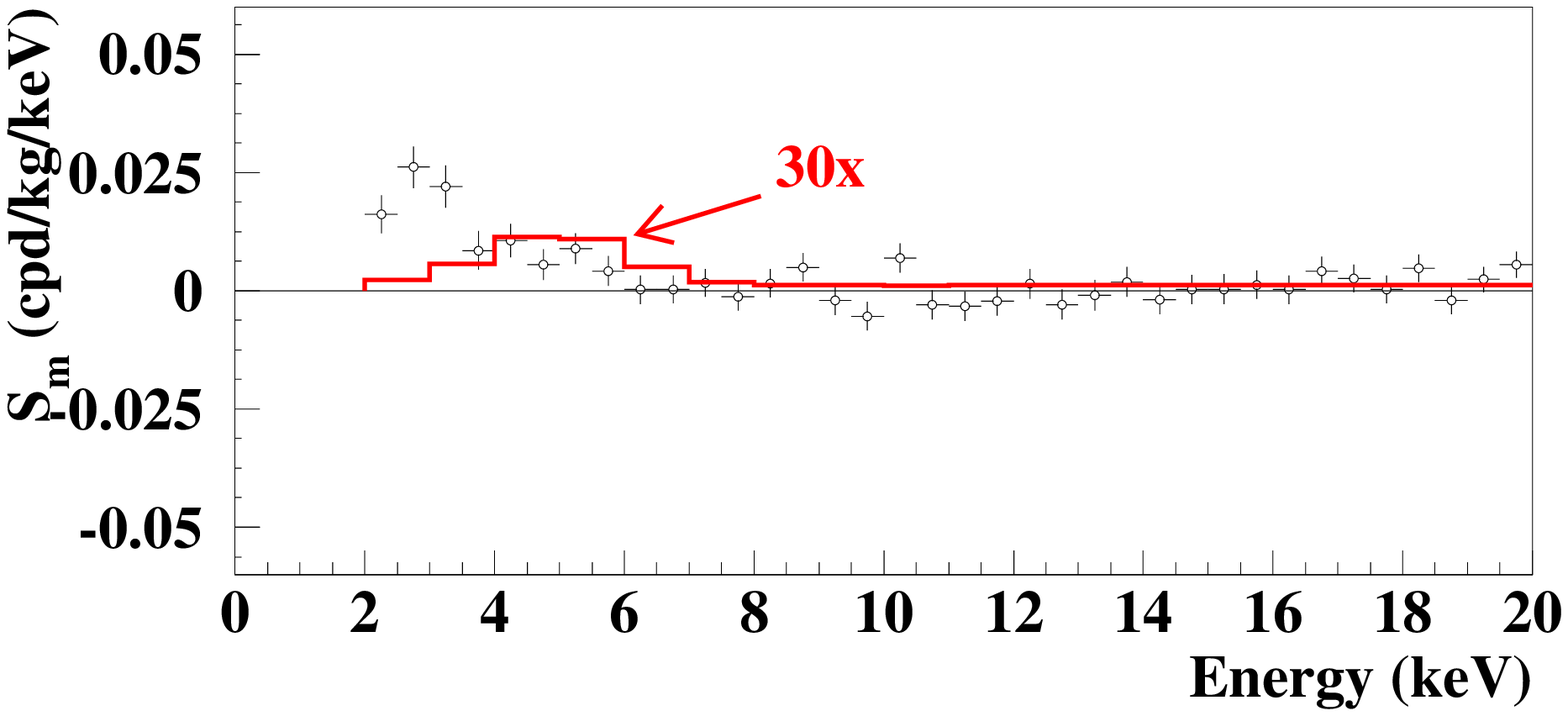}
\end{center}
\vspace{-0.8cm}
\caption{{\it Top -} Data points: cumulative low-energy
distribution of the {\it single-hit} scintillation events
measured by DAMA/LIBRA \cite{perflibra} above the 2 keV energy threshold of the experiment.
Histogram (color online): maximum expected counting rate from $^{128}$I decays corresponding 
to the measured upper limit on $^{128}$I activity in the NaI(Tl) detectors: $<$14.8 $\mu$Bq/kg 
(90\% C.L.); see the data in ref. \cite{papep} and the text. {\it Bottom -} Data points:
the DAMA measured modulation amplitude as a function of the energy. Histogram (color online): maximum 
expected modulation amplitude multiplied by a factor 30 as a function of the energy from $^{128}$I decays 
corresponding to the measured upper limit on $^{128}$I activity given above and  
assuming the hypothetical case that the neutron flux 
had a 10\% time modulation with the same phase, period as a DM signal. 
Therefore, the contribution from $^{128}$I is negligible and cannot mimic the
S$_m$ behaviour; in addition, it is worth 
noting that $^{128}$I never could mimic the DM annual 
modulation signature since some of its peculiarities would fail (e.g. $^{128}$I
would induce modulation also in other energy regions, which is not observed). See text.} 
\label{128I}
\vspace{-0.4cm}
\end{figure}
Moreover, the data collected by DAMA/LIBRA allow the determination
of the possible presence of $^{128}$I in the detectors.
In fact, neutrons would generate $^{128}$I homogeneously distributed in the
NaI(Tl) detectors; therefore studying the characteristic radiation of
$^{128}$I decay and comparing it with the experimental data, one can obtain
the possible $^{128}$I concentration. The most sensitive way to perform such a
measurement is to study the possible presence
of the 32 keV peak ($K$-shell contribution) in the region around 30 keV.
This has already been done by DAMA -- for other purposes -- in ref. \cite{papep},
where there is also no evidence of such a peak in the DAMA/LIBRA data;
hence an upper limit on the area of a peak around 32 keV can be derived to
be: 0.074 cpd/kg (90\% CL) \cite{papep}. Considering the branching ratio 
of the process and the related efficiency, one can obtain a limit on possible 
activity of $^{128}$I: $<14.8 \mu$Bq/kg (90\% CL). This upper limit allows to derive the  
maximum expected counting rate from $^{128}$I (see  Fig. \ref{128I}--{\it Top}), 
showing that in every case its 
contribution -- if any -- is negligible. The contribution is also negligible
even in the hypothetical case that the neutron flux
had a 10\% time modulation with the same phase, period as a DM signal;
in fact, even in such a case 
the contribution to S$_m$ is $<3\times10^{-4}$ cpd/kg/keV at low energy (see  Fig. \ref{128I}--{\it Bottom}),
that is $<2\%$ of the DAMA observed modulation amplitudes.
Therefore, for all the given arguments (and just one of them is enough), 
no role is played by $^{128}$I.

\section{Conclusions}

As regards the corollary investigation on the nature of the DM candidate particle(s) 
and related astrophysical, nuclear and particle physics
scenarios, it has been shown that the obtained model independent evidence 
can be compatible with a wide set of possibilities as discussed e.g. in refs.  
\cite{RNC,ijmd,epj06,ijma07,inel,ijma,wimpele,ldm,chan},
in the Appendix of ref. \cite{modlibra} and 
in literature (as e.g. \cite{fen}, etc.);
other possibilities are open.
Moreover, as regards possible comparisons with direct, 
indirect and accelerators 
activities see e.g. discussions in 
\cite{modlibra2,modlibra,RNC,paperliq,taupnoz,collar}, 
etc..
Here we just recall that no other experiment
exists, whose result can be directly compared in a model-independent
way with those by DAMA/NaI and DAMA/LIBRA.

In conclusion, the six annual cycles of DAMA/LIBRA have further confirmed the
peculiar annual modulation of the {\it single-hit} events in the (2--6) keV energy region.
The total exposure by the former DAMA/NaI and the present DAMA/LIBRA is
1.17 ton $\times$ yr and the confidence level for the observed
effect is cumulatively about 9 $\sigma$ CL. The data satisfy all the many
requirements of the signature and no systematics or
side reactions able to mimic it (that is, able to account
for the measured modulation amplitude and to simultaneously
satisfy all the peculiarities of the signature) have been found
or suggested by anyone over more than a decade. 

In near future new PMTs 
with higher quantum efficiency will be installed in order to lower the 
2  keV energy threshold, increasing the experimental sensitivity and improving  
the 
corollary information on the nature of the DM candidate particle(s) and on the various related 
astrophysical, nuclear and particle physics scenarios.
Moreover, it will also allow the investigation
of other DM features, of
second order effects and of several rare processes other than DM.
A third generation R\&D effort towards a possible NaI(Tl) ton
set-up, DAMA proposed in 1996, has been funded by I.N.F.N. and is in progress.

\end{document}